\documentstyle[12pt,epsfig]{article}

\newcommand{\half}{{\textstyle\frac{1}{2}}}
\newcommand{\ee}{e^+e^-}
\newcommand{\as}{\alpha_s}

\newcommand{\asPT}{\alpha_s^{\mbox{\scriptsize PT}}}
\newcommand{\eps}{\epsilon}
\newcommand{\beq}{\begin{equation}}
\newcommand{\eeq}{\end{equation}}
\newcommand{\cl}[1]{{\cal #1}}
\newcommand{\rf}[1]{(\ref{#1})}
\newcommand{\sect}[1]{\section{#1}\setcounter{equation}{0}}
\newcommand{\nln}{\nonumber\\}
\newcommand{\qtot}{q_{\mbox{\scriptsize tot}}}
\newcommand{\qwtd}{q_{\mbox{\scriptsize c-w}}}
\newcommand{\Dwtd}{D_{\mbox{\scriptsize c-w}}^h}
\renewcommand{\theequation}{\arabic{section}.\arabic{equation}}

\newcommand{\prlett}[1]{{\it Phys.~Rev.~Lett.~\bf #1}}
\newcommand{\nphysb}[1]{{\it Nucl.~Phys.~\bf B #1}}
\newcommand{\plettb}[1]{{\it Phys.~Lett.~\bf B #1}}
\newcommand{\phrevd}[1]{{\it Phys.~Rev.~\bf D #1}}
\newcommand{\jinrrc}[1]{{\it JINR~Rapid~Comm.~\bf #1}}
\newcommand{\npproc}[1]{{\it Nucl.~Phys.~Proc.~Suppl.~\bf #1}}
\newcommand{\jhephy}[1]{{\it JHEP~\bf #1}}
\newcommand{\ephysj}[1]{{\it Eur.~Phys.~J.~\bf #1}}
\newcommand{\zphysc}[1]{{\it Z.~Phys.~\bf C #1}}
\newcommand{\sovphy}[1]{{\it Sov.~Phys.~JETP~\bf #1}}
\newcommand{\mpleta}[1]{{\it Mod.~Phys.~Lett.~\bf A #1}}

\begin{document}
\begin{titlepage}
\begin{flushright}
Cavendish-HEP-98/16\\
hep-ph/9812251\\
December 1998
\end{flushright}              
\vspace*{\fill}
\begin{center}
{\Large \bf Power Corrections to Fragmentation Functions\\[1ex]
in Flavour-Singlet Deep Inelastic Scattering\footnote{Research supported by the U.K. Particle Physics and Astronomy Research Council.}}
\end{center}
\par \vskip 5mm
\begin{center}
        G.E.~Smye\\
        Cavendish Laboratory, University of Cambridge,\\
        Madingley Road, Cambridge CB3 0HE, U.K.\\
\end{center}
\par \vskip 2mm
\begin{center} {\large \bf Abstract} \end{center}
\begin{quote}
We investigate the power-suppressed corrections to fragmentation functions in flavour singlet deep inelastic lepton scattering, to complement the previous results for the non-singlet contribution. Our method is a dispersive approach based on an analysis of Feynman graphs containing massive gluons. As in non-singlet deep inelastic scattering we find that the leading corrections are proportional to $1/Q^2$.
\end{quote}
\vspace*{\fill}
\end{titlepage}

\sect{Introduction}
The study of fragmentation functions in deep inelastic lepton scattering (DIS) has received a great impetus from the increasing quantity and kinematic range of the HERA data \cite{expfrgfns}. While these functions cannot be calculated perturbatively in QCD, their asymptotic scaling violations (logarithmic $Q^2$ dependence) can be calculated, enabling measurements of the strong coupling constant $\as$. In addition there are power-suppressed (higher-twist) contributions to the scaling dependence, which
need to be estimated in order to make use
of the wide $Q^2$ coverage of HERA.

Recently so-called `renormalon' or `dispersive' methods of
estimating power-suppressed terms have been formulated: for descriptions see \cite{renlons,dmw,asmodel}. Such approaches have been applied with some success to $\ee$
fragmentation functions \cite{eefrag} and event shape variables \cite{eeevsh},
and to flavour non-singlet DIS structure functions \cite{dmw,dwdisstr,disstr},
fragmentation functions \cite{disfrg} and event shape variables \cite{disev}.
Note that special care has to be taken when considering non-inclusive
observables such as fragmentation functions and event shape variables
\cite{milan}. There are also renormalon model results for photon-photon
scattering \cite{hautmann}
and structure functions in flavour singlet DIS \cite{smye,stein}.

In the present paper we extend the dispersive method to fragmentation functions in flavour singlet DIS, using many of the same techniques as in \cite{smye}. As in non-singlet DIS and $\ee$ annihilation, the predicted leading power corrections to these quantities are proportional to $1/Q^2$, but their functional forms are different. In particular there are contributions proportional to $(\log Q^2)/Q^2$, which are not found in the non-singlet case. While the magnitude of the correction is not known, the hypothesis that it is related to a universal low-energy strong coupling implies that it is given by universal non-perturbative parameters.

In the following section we review the approach of references \cite{dmw,smye}. Section \ref{fragfns} presents the standard leading-order perturbative treatment of DIS fragmentation functions. In section \ref{calc} we estimate the power-suppressed corrections using the method outlined in section \ref{dispsec}. Our results are summarized briefly in section \ref{concl}.

\sect{The Dispersive Approach to Power Corrections}
\label{dispsec}
We assume that the QCD running coupling $\alpha_s(k^2)$ can be defined for all positive $k^2$, and that apart from a branch cut along the negative real axis there are no singularities in the complex plane. It follows that we may write the formal dispersion relation:
\beq
\label{disprel}
\as(k^2) = - \int_0^\infty\frac{d\mu^2}{\mu^2+k^2}\rho_s(\mu^2)\; ,
\eeq
where the `spectral function' $\rho_s$ represents the discontinuity across the cut:
\beq
\rho_s(\mu^2)=\frac{1}{2\pi i}\biggl\{\as(\mu^2 e^{i\pi})-\as(\mu^2 e^{-i\pi})\biggr\} = \frac{1}{2\pi i}\mbox{Disc }\as(-\mu^2)\; .
\eeq

Non-perturbative effects at long distances are expected to give rise to a non-perturbative modification to the perturbatively-calculated strong coupling at low scales, $\delta\as(\mu^2) = \as(\mu^2) - \asPT(\mu^2)$, $\asPT(\mu^2)$ being the perturbatively-calculated running coupling \cite{dmw}. This modification gives rise to power-behaved corrections to QCD observables. (Note that here $\asPT(\mu^2)$ refers only to the next-to-leading order perturbatively-calculated running coupling, and so is itself well-behaved down to low scales, without any divergences: the Landau pole appears when we include an arbitrary number of loop insertions in the propagator. Hence $\delta\as(\mu^2)$ is a well-defined quantity for all positive $\mu^2$.)

We now consider the calculation of some observable $F$ in an improved one-loop approximation which takes into account one-gluon contributions plus those higher-order terms that lead to the running of $\as$. We obtain the characteristic function $\cl{F}(\mu^2/Q^2)$, which is a one-loop evaluation of $F$ but with the gluon mass set equal to $\mu$. We may then write \cite{dmw}
\beq
\label{Feq}
F = \int_0^\infty \frac{d\mu^2}{\mu^2}\,\rho_s(\mu^2)\,\left[\cl{F}(\mu^2/Q^2)-\cl{F}(0)\right]\;.
\eeq
The behaviour of $\cl{F}(\eps)$ at small $\eps$ generates a correction to $F$, which we denote $\delta F$. Terms of $\cl{F}(\eps)$ which are analytic at $\eps=0$ do not contribute to $\delta F$ \cite{dmw}; but non-analytic terms at small $\eps$ do contribute, the relevant terms being:
\beq
\cl{F}\sim a_1 \frac{C_F}{2\pi}\sqrt{\eps}
\qquad\Longrightarrow\qquad
\delta F = -\frac{a_1}{\pi}\frac{\cl{A}_1}{Q}\;,
\eeq
and
\beq
\cl{F}\sim a_2 \frac{C_F}{2\pi}\eps\log\eps
\qquad\Longrightarrow\qquad
\delta F = a_2\frac{\cl{A}_2}{Q^2}\;,
\eeq
where
\beq
\label{Aqdef}
\cl{A}_q \equiv \frac{C_F}{2\pi}\int_0^\infty \frac{d\mu^2}{\mu^2}\,\mu^q\,\delta\as(\mu^2)\;.
\eeq

Studies of the non-singlet contribution to DIS structure functions suggest that $\cl{A}_2 \simeq 0.2 \mbox{GeV}^2$ \cite{dwdisstr}.

This is the formulation of the dispersive approach to power behaved corrections used where there is a single gluon propagator. However, the calculation which follows in section \ref{calc} involves two gluons, and so the above argument needs to be generalised.

Each gluon has an associated dispersive variable, so we obtain a characteristic function in both of these, $\cl{F}(\eps_1,\eps_2)$. Following the above argument, we require those terms of $\cl{F}$ which are non-analytic at 0 in both arguments. However, in the case where there are two internal gluons constrained to have the same 4-momentum $k$, we can simplify this to require only one dispersive variable. By defining $\rho = - k^2 / Q^2$, we see that the dependence of the characteristic function $\cl{F}$ on $\eps_1$ and $\eps_2$ is given by
\beq
\cl{F}(\eps_1,\eps_2) = \int\frac{d\rho\,f(\rho)}{(\rho+\eps_1)(\rho+\eps_2)}\;,
\eeq
where the integration limits and the function $f$ depend on the particular calculation. This may be expressed in partial fractions in the form
\beq
\label{Fdecomp}
\cl{F}(\eps_1,\eps_2) = \frac{\eps_1\hat{\cl{F}}(\eps_1)-\eps_2\hat{\cl{F}}(\eps_2)}{\eps_1-\eps_2}\;,
\eeq
where
\beq
\hat{\cl{F}}(\eps) = -\frac{1}{\eps}\int\frac{d\rho\,f(\rho)}{\rho+\eps}\;.
\eeq

Now the generalisation of the one-gluon result \rf{Feq} to the multiple-gluon case (which can easily be seen from a generalisation of the argument in \cite{dmw}) is
\beq
F = \int_0^\infty\frac{d\mu_1^2}{\mu_1^2}\frac{d\mu_2^2}{\mu_2^2}\rho_s(\mu_1^2)\rho_s(\mu_2^2)[\cl{F}(\mu_1^2/Q^2,\mu_2^2/Q^2)-\cl{F}(\mu_1^2/Q^2,0)-\cl{F}(0,\mu_2^2/Q^2)+\cl{F}(0,0)]\;.
\eeq
Substituting \rf{Fdecomp} into this gives, as shown in \cite{smye}, that the relevant non-analytic terms in $\hat{\cl{F}}(\eps)$ are:
\beq
\hat{\cl{F}}\sim a_1 \eps\log\eps
\qquad\Longrightarrow\qquad
\delta F = a_1 \frac{D_1}{Q^2}\;,
\eeq
and
\beq
\hat{\cl{F}}\sim \half a_2 \eps\log^2\eps
\qquad\Longrightarrow\qquad
\delta F = a_2\frac{D_1}{Q^2}\log\frac{D_2}{Q^2}\;,
\eeq
where $D_1$ and $D_2$ are defined by:
\begin{eqnarray}
\label{D1def}
D_1 &\equiv& \int_0^\infty \frac{d\mu^2}{\mu^2}\,\mu^2\,\left(2\as(\mu^2)\delta\as(\mu^2)-[\delta\as(\mu^2)]^2\right)\;,\\
\label{D2def}
\log D_2 &\equiv& \frac{1}{D_1}\int_0^\infty \frac{d\mu^2}{\mu^2}\,\mu^2\log\mu^2\,\left(2\as(\mu^2)\delta\as(\mu^2)-[\delta\as(\mu^2)]^2\right)\;.
\end{eqnarray}

While we expect the form of $\as(\mu^2)$, and hence $D_1$ and $D_2$, to be universal, we have as yet no numerical values for them, (unlike for $\cl{A}_2$, which is defined by \rf{Aqdef} with $q=2$). It will be necessary therefore to extract values for $D_1$ and $D_2$, either from experimental results or from some model of the form of $\alpha_s(\mu^2)$ (of which various models have been proposed \cite{asmodel}).

\sect{DIS Fragmentation Functions}
\label{fragfns}
We consider the deep inelastic scattering of a lepton of momentum $l$ from a nucleon of momentum $P$, with transfer of a virtual photon of momentum $q$. The standard variables for describing DIS are $Q = \sqrt{-q^2}$ ($Q \ge 0$), the Bjorken variable $x = Q^2 / 2 P\cdot q$ ($0\le x\le 1$), and $y = P\cdot q / P\cdot l\simeq Q^2/x s$, where $s$ is the total c.m. energy squared, ($0\le y\le 1$).

\begin{figure}[ht]
\begin{center}
\epsfig{file=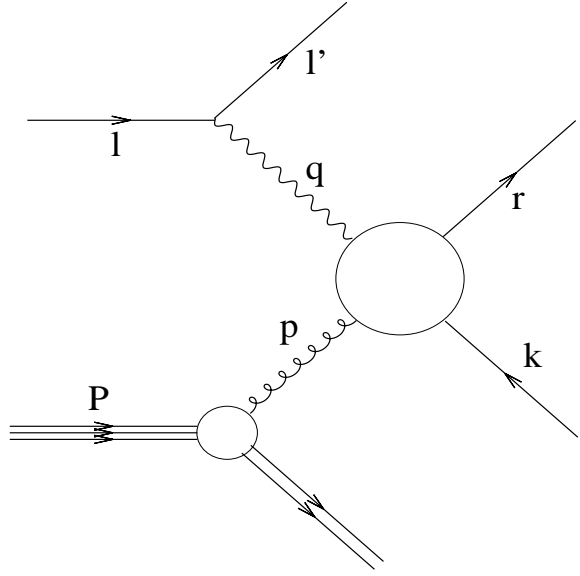, height=3.0in, width=3.0in}
\caption{\label{singfig}Flavour singlet contribution to deep inelastic scattering}
\end{center}
\end{figure}

We wish to study the fragmentation function $F^h(z)$, defined by
\begin{equation}
\label{fragdef}
F^h(z;x,Q^2) = \frac{d^3\sigma^h}{dxdQ^2dz}\left/\frac{d^2\sigma}{dxdQ^2}\right.\; ,
\end{equation}
for a given hadron species $h$, as a function of the variable $z = 2 p_h\cdot q/q^2$, where $p_h$ is the momentum of the resultant hadron. Since the fragmentation products of the remnant of the nucleon are expected to be travelling in directions close to that of the incoming nucleon, i.e. in the `remnant hemisphere' $p_z \le 0$, we consider only hadrons produced in the `current hemisphere' $p_z \ge 0$. Such hadrons are expected to be fragmentation products of the scattered parton. Thus $z$ takes values $0\le z\le 1$.

The differential cross sections in (\ref{fragdef}) can be decomposed into
\begin{eqnarray}
\frac{d^3\sigma^h}{dxdQ^2dz} &=& \frac{2\pi\alpha^2}{Q^4}\Bigl\{[1+(1-y)^2]F_T^h(x,z) + 2(1-y)F_L^h(x,z)\Bigr\}\\
\frac{d^2\sigma}{dxdQ^2} &=& \frac{2\pi\alpha^2}{Q^4}\Bigl\{[1+(1-y)^2]F_T(x) + 2(1-y)F_L(x)\Bigr\}\; ,
\end{eqnarray}
where $F_T$ and $F_L$ are the transverse and longitudinal structure functions and $F_T^h$ and $F_L^h$ are their generalisations to include fragmentation. They are related to $F_1$ and $F_2$ by $F_T(x) = 2F_1(x)$ and $F_L(x) = F_2(x)/x - 2F_1(x)$.

The parton model assumptions are that $p = xP/\xi$ ($x\le\xi\le 1$) and $p_h = zp_p/\zeta$ ($z\le\zeta\le 1$), where $p_p$ is the momentum of the fragmenting parton (i.e. for the diagram in figure \ref{singfig}, $p_p = r$ for quark fragmentation and $p_p = k$ for antiquark fragmentation). Thus we get
\begin{eqnarray}
F_i^h(x,z) &=& \sum_{\parbox{0.5in}{\scriptsize\raggedright incoming partons $j$}}\sum_{\parbox{0.5in}{\scriptsize\raggedright outgoing partons $k$}}\int_x^1\frac{d\xi}{\xi}\int_z^1\frac{d\zeta}{\zeta}j(x/\xi)D_k^h(z/\zeta)F_{i,jk}(\xi,\zeta)\\
F_i(x) &=& \sum_{\parbox{0.5in}{\scriptsize\raggedright incoming partons $j$}}\int_x^1\frac{d\xi}{\xi}j(x/\xi)F_{i,j}(\xi)
\end{eqnarray}
where $i=T$ or $L$, $j(x/\xi)$ is the parton distribution function for the parton $j$, $D_k^h(z/\zeta)$ is the fragmentation function for the parton $k$ into the hadron $h$, and $F_{i,jk}(\xi,\zeta)$ and $F_{i,j}(\xi)$ are the (generalised) parton level structure functions. Note that all the functions in the above equations have a scaling violation or $Q^2$ dependence which is not shown explicitly. (For simplicity we are neglecting any contribution from weak interactions, i.e. $\mbox{Z}^0$ or $\mbox{W}^\pm$ exchange.)

In the parton model, to order $\as^0$, we have
\begin{eqnarray}
\label{ord0t}
F_T^h(x,z) &=& \sum_q e_q^2 [q(x)D_q^h(z)+\bar{q}(x)D_{\bar{q}}^h(z)]\\
F_L^h(x,z) &=& 0\frac{}{}\\
F_T(x) &=& \sum_q e_q^2 [q(x)+\bar q(x)]\\
\label{ord0b}
F_L(x) &=& 0
\end{eqnarray}
where $q(x)$ and $\bar{q}(x)$ are the quark and antiquark distributions in the target nucleon.

There are $\cl{O}(\as)$ contributions arising both from real and virtual gluon emission, as described in \cite{disfrg}. Here we concentrate on the singlet contribution, i.e.~the contribution from the gluon content of the target hadron, as shown in figure \ref{singfig}. There are no virtual corrections to the singlet contribution at this order.

\sect{Power Corrections in Flavour Singlet DIS}
\label{calc}
In the normal perturbative treatment of DIS, the asymptotic freedom of QCD enables us to treat the initial state partons as free particles confined within the nucleon; and so in a singlet calculation we would start from a free gluon and convolute the perturbative result with the gluon distribution function $g(x)$. We do not know how to do this in a calculation of power corrections, since the models we use consider modifications to the gluon propagator (loop insertions in the renormalon model, or, equivalently, a `mass' in the dispersive approach).
Let us therefore perform the calculation by considering our initial state gluon to be radiated from a fermionic parton. We may then try to recover the singlet contribution to the power corrections by deconvoluting the result with the quark to gluon splitting function, as performed in \cite{stein}, or we may leave the result as it is and interpret it as a second-order non-singlet contribution. We might hope that these two interpretations would give similar predictions for power-suppressed corrections.

\begin{figure}[ht]
\begin{center}
\epsfig{file=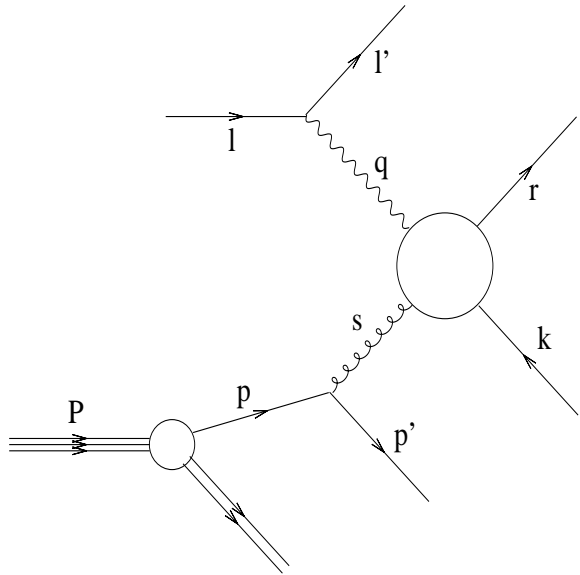, height=3.0in, width=3.0in}
\caption{\label{diags}Diagram generating flavour singlet contribution}
\end{center}
\end{figure}

Consider the contribution to DIS from the diagrams in figure \ref{diags}. It is convenient to work in the {\it Breit frame of reference} \cite{bframe}, which is the rest-frame of $2xP+q$. In this frame the momentum transfer $q$ is purely spacelike, and we choose to align it along the $+z$ axis.

The momentum of the initial state parton is $p = xP/\xi$, $(x\le\xi\le 1)$; and let us introduce the variables $\rho = - s^2 / Q^2$, $\eta = P\cdot r/P\cdot q$, $\bar{\eta} = P\cdot k/P\cdot q$ , $\chi$ the azimuthal angle between $r$ and $s$, and $\gamma$ the azimuthal angle between $r$ and $k$. There is also an overall azimuthal angle $\phi$.

In the Breit frame the kinematics are given by:
\begin{eqnarray}
P &=& \half Q (1/x,0,0,-1/x) \\
p &=& \half Q (1/\xi,0,0,-1/\xi) \\
q &=& \half Q (0,0,0,2) \\
s &=& \half Q (s_0,s_\perp\cos\chi,s_\perp\sin\chi,s_3) \\
r &=& \half Q (z_0,z_\perp,0,z_3) \\
k &=& \half Q (\bar{z}_0,\bar{z}_\perp\cos\gamma,\bar{z}_\perp\sin\gamma,\bar{z}_3)\;.
\end{eqnarray}

The definitions of $\rho$, $\eta$ and $\bar{\eta}$ along with the on-shell conditions for the outgoing particles require that
\begin{eqnarray}
s_0 = \frac{1}{\xi}-\rho\xi-\frac{s_\perp^2}{4\rho\xi}\hspace{0.3in} & &
s_3 = -\frac{1}{\xi}-\rho\xi+\frac{s_\perp^2}{4\rho\xi} \\
z_0 = \eta+\frac{z_\perp^2}{4\eta}\hspace{0.75in} & &
z_3 = \eta-\frac{z_\perp^2}{4\eta} \\
\bar{z}_0 = \bar{\eta}+\frac{\bar{z}_\perp^2}{4\bar{\eta}}\hspace{0.75in} & &
\bar{z}_3 = \bar{\eta}-\frac{\bar{z}_\perp^2}{4\bar{\eta}}\;.
\end{eqnarray}

Conservation of the 0th and 3rd components of 4-momentum give the conditions
\begin{eqnarray}
\eta + \bar{\eta} + \rho\xi &=& 1 \\
\frac{z_\perp^2}{4\eta}+\frac{\bar{z}_\perp^2}{4\bar{\eta}}+\frac{s_\perp^2}{4\rho\xi} &=& \frac{1-\xi}{\xi}\;,
\end{eqnarray}
while conservation of transverse momentum requires that $s_\perp$, $z_\perp$ and $\bar{z}_\perp$ satisfy the triangle inequalities
\begin{eqnarray}
\vert z_\perp-s_\perp\vert &\le& \bar{z}_\perp \\
\vert z_\perp-\bar{z}_\perp\vert &\le& s_\perp\;.
\end{eqnarray}

Two variables, $\alpha$ and $\beta$, are required to parametrise the permitted values of $s_\perp$, $z_\perp$ and $\bar{z}_\perp$. Let us choose to write:
\begin{eqnarray}
z_\perp^2 &=& \frac{4\alpha\eta(1-\eta)(1-\xi)}{\xi} \\
\bar{z}_\perp^2 &=& \frac{4\bar{\eta}(1-\xi)}{(1-\eta)\xi}\left[(1-\alpha)\rho\xi+\alpha\eta\bar{\eta}-2\cos\beta\sqrt{\alpha(1-\alpha)\eta\bar{\eta}\rho\xi}\right] \\
s_\perp^2 &=& \frac{4\rho(1-\xi)}{1-\eta}\left[(1-\alpha)\bar{\eta}+\alpha\eta\rho\xi+2\cos\beta\sqrt{\alpha(1-\alpha)\eta\bar{\eta}\rho\xi}\right]\;.
\end{eqnarray}
We could equally have used $\bar{\alpha}$ and $\bar{\beta}$, defined similarly but with the outgoing quark and antiquark interchanged. In particular, if $\bar{z}_\perp^2 = 4\bar{\alpha}\bar{\eta}(1-\bar{\eta})(1-\xi)/\xi$, then
\beq
\bar{\alpha} = \frac{1}{(1-\eta)(1-\bar{\eta})}\left[(1-\alpha)\rho\xi+\alpha\eta\bar{\eta}-2\cos\beta\sqrt{\alpha(1-\alpha)\eta\bar{\eta}\rho\xi}\right]\;.
\eeq

Given $s_\perp$, $z_\perp$ and $\bar{z}_\perp$, the angles $\chi$ and $\gamma$ are determined up to a sign. We may then choose $\alpha$, $\beta$, $\eta$ and $\rho$ as the independent variables, with phase space
\begin{equation}
0\le\alpha\le 1,\;0\le\beta\le\pi,\;0\le\eta\le 1,\;0\le\rho\le (1-\eta)/\xi .
\end{equation}

To estimate power corrections to perturbative calculations, we must perform the calculations as though the gluons had small non-zero masses $\mu_1^2 = \epsilon_1 Q^2$ and $\mu_2^2 = \epsilon_2 Q^2$. Using the machinery of section \ref{dispsec} we may write
\begin{equation}
{\cal F}_i(x,z;\epsilon_1,\epsilon_2) = \frac{\epsilon_1\hat{\cal F}_i(x,z;\epsilon_1)-\epsilon_2\hat{\cal F}_i(x,z;\epsilon_2)}{\epsilon_1-\epsilon_2}\;,
\end{equation}
where, by the symmetry between the outgoing quark and antiquark,
\begin{equation}
\hat{\cal F}_i(x,z;\epsilon) = \frac{\alpha_s^2 T_R C_F}{(2\pi)^2}\sum_q\sum_{q^\prime}e_{q^\prime}^2\int_x^1\frac{d\xi}{\xi}\int_z^1\frac{d\zeta}{\zeta}C_i(\xi,\zeta;\epsilon)q(x/\xi)\left[D_{q^\prime}^h(z/\zeta)+D_{\bar{q}^\prime}^h(z/\zeta)\right]\;,
\end{equation}
and $\zeta$ is given in terms of the other variables by
\beq
\zeta = \frac{2r\cdot q}{q^2} = \frac{\xi+\alpha(1-\xi)}{\xi}\eta-\frac{\alpha(1-\xi)}{\xi}\;.
\eeq

The quantities $C_i$ are the integrated perturbative matrix elements with modified gluon propagators.

To integrate the matrix elements we apply the operator
\begin{equation}
\int\frac{d^3\b{r}}{(2\pi)^3 2r^0}\frac{d^3\b{k}}{(2\pi)^3 2k^0}\frac{d^3\b{p}^\prime}{(2\pi)^3 2p^{\prime 0}}\:(2\pi)^4\delta^4(p+q-k-r-p^\prime)\;.
\end{equation}
Integrating away $\b{p}^\prime$ and making substitutions for $\b{r}$ and $\b{k}$ gives
\begin{equation}
\frac{Q^2}{32(2\pi)^5}\int\frac{z_\perp dz_\perp d\phi d\eta}{\eta}\frac{\bar{z}_\perp d\bar{z}_\perp d\gamma d\bar{\eta}}{\bar{\eta}}\:\frac{\delta\left(A+\sqrt{B+2z_\perp\bar{z}_\perp\cos\gamma}\right)}{\sqrt{B+2z_\perp\bar{z}_\perp\cos\gamma}}\;,
\end{equation}
where $A$ and $B$ do not depend on $\gamma$.

Next we integrate over $\gamma$. There are two values satisfying the delta-function condition, and they differ by a sign. This gives
\begin{equation}
\frac{Q^2}{32(2\pi)^5}\int\frac{d\eta}{\eta}\frac{d\bar{\eta}}{\bar{\eta}}\frac{dz_\perp^2 d\bar{z}_\perp^2}{2z_\perp\bar{z}_\perp\vert\sin\gamma\vert}d\phi\;.
\end{equation}

Applying the parametrisation in terms of $\alpha$ and $\beta$, we find that
\begin{eqnarray}
\frac{\partial(z_\perp^2,\bar{z}_\perp^2)}{\partial(\alpha,\beta)} &=& 32(1-\xi)^2\sin\beta\sqrt{\alpha(1-\alpha)\eta^3\bar{\eta}^3\rho/\xi^3} \\
2z_\perp\bar{z}_\perp\vert\sin\gamma\vert &=& 8(1-\xi)\sin\beta\sqrt{\alpha(1-\alpha)\eta\bar{\eta}\rho/\xi}\;.
\end{eqnarray}
Therefore the integral operator is
\begin{equation}
\frac{Q^2(1-\xi)}{8(2\pi)^5}\int d\rho d\eta d\alpha d\beta d\phi\;.
\end{equation}

We are interested to find the differential coefficient functions as a function of $\zeta$. In the current hemisphere ($\zeta>0$) there is a singularity at $\zeta=1$, associated with the antiquark becoming soft or collinear with the proton. This divergence must be investigated separately. The divergences associated with the quark becoming soft or collinear occur when $\zeta\le 0$, and so are not seen in the current hemisphere.

For $0<\zeta<1$, then we have
\begin{equation}
C_i(\xi,\zeta;\epsilon) = -\frac{1}{\epsilon}\frac{Q^2(1-\xi)}{8(2\pi)^5}\int_0^1d\alpha\int_0^{\frac{1-\zeta}{\xi+\alpha(1-\xi)}}d\rho\int_0^\pi d\beta\int_0^{2\pi}d\phi\frac{\xi}{\xi+\alpha(1-\xi)}\frac{f}{\rho+\epsilon}\;,
\end{equation}
where the function $f(\xi,\zeta,\alpha,\beta,\phi,\rho)$ is obtained from the appropriate matrix elements, excluding the denominators of the gluon propagators. We calculate $C_T$ and $C_L$ using projections of the hadronic tensor representing the QCD part of the process. Since we wrote the kinematics without $\phi$-dependence, the projections obtain $\phi$-dependence to compensate. We find that the integration over $\phi$ is trivial for $C_L$ and elementary for $C_T$.

The part of this that is non-analytic as $\epsilon\rightarrow 0$ is the contribution to the integral near $\rho = 0$. Therefore we proceed by expressing $f$ as a series expansion in $\rho^{\frac{1}{2}}$ about 0, up to ${\cal O}(\rho^2)$. The integral over $\beta$ is then straightforward, being a sum of terms of the form $\mbox{const.}\cos^n\beta$, and upon performing this integral the terms in non-integer powers of $\rho$ vanish. Then note that the non-analytic parts of the integrals over $\rho$ are given by:
\begin{eqnarray}
\int_0^c\frac{\rho^n}{\rho+\epsilon}d\rho &\to& (-1)^{n-1}\epsilon^n\log\epsilon\;, \\
\int_0^c\frac{\rho^n\log\rho}{\rho+\epsilon}d\rho &\to& {\textstyle\frac{1}{2}}(-1)^{n-1}\epsilon^n\log^2\epsilon\;.
\end{eqnarray}
The integral over $\alpha$ is straightforward.

To study the divergence at $\zeta=1$, consider, for some small $\Delta$ and some arbitrary function $g(\zeta)$,
\begin{eqnarray}
\lefteqn{\int_{1-\Delta}^1 d\zeta C_i(\xi,\zeta;\eps)g(\zeta)}\nln
&=& -\frac{1}{\epsilon}\frac{Q^2(1-\xi)}{8(2\pi)^5}\int_0^1d\alpha\int_0^{\frac{\Delta}{\xi+\alpha(1-\xi)}}d\rho\int_{1-\frac{\xi\Delta}{\xi+\alpha(1-\xi)}}^{1-\rho\xi}d\eta\int_0^\pi d\beta\int_0^{2\pi}d\phi\frac{f}{\rho+\epsilon}g(\zeta)
\end{eqnarray}

In order to produce valid series expansions in $\rho$, it transpires that we have to divide the $\eta$ integral into two regions:
\begin{enumerate}
\item $1-\frac{\xi\Delta}{\xi+\alpha(1-\xi)}\le\eta\le 1-\rho\xi-\rho/\kappa$, and
\item $1-\rho\xi-\rho/\kappa\le\eta\le 1-\rho\xi$,
\end{enumerate}
where $\kappa$ is some small arbitrary quantity on which the final answer should not depend. (In practice we perform the calculation in the limit of small $\kappa$.)

For region (i) we may directly expand $f$ in $\rho$ without any problems; and since $\Delta$ is small we can expand $g(\zeta)$ as
\beq
g(\zeta) = g(1)-\frac{\xi+\alpha(1-\xi)}{\xi}(1-\eta)g^\prime(1)+\cl{O}(\Delta^2)\;.
\eeq

Note that because of the limits we have chosen for the $\eta$ integral, contributions up to ${\cal O}(\rho^2)$ will also arise from all higher terms in the series expansion of $f$. These contributions vanish as $\kappa\rightarrow 0$, so the calculation performed in this limit is valid.

For region (ii) such a naive power series becomes invalid. We use the symmetry between the outgoing quark and antiquark to see that $f(\xi,\eta,\alpha,\beta,\phi,\rho) = f(\xi,\bar{\eta},\bar{\alpha},\bar{\beta},\phi,\rho)$ and that the integration measure $d\eta d\alpha d\beta = d\bar{\eta}d\bar{\alpha}d\bar{\beta}$. We may therefore write the required integral as:
\beq
-\frac{1}{\epsilon}\frac{Q^2(1-\xi)}{8(2\pi)^5}\int_0^1d\alpha\int_0^{\frac{\Delta}{\xi+\alpha(1-\xi)}}d\rho\int_0^{\rho/\kappa}d\eta\int_0^\pi d\beta\int_0^{2\pi}d\phi\frac{f}{\rho+\epsilon}g(\bar{\zeta})\;.
\eeq
We may now write $\eta = \lambda\rho$ and then expand $f$ as before. We may also expand $g$ as:
\beq
g(\bar{\zeta}) = g(1)-\frac{\xi+\bar{\alpha}(1-\xi)}{\xi}(1-\bar{\eta})g^\prime(1)+\cl{O}(\Delta^2)\;.
\eeq
In contrast to the previous situations there are now no terms in non-integer powers of $\rho$. The integration over $\beta$ may be performed by writing $t = \tan(\beta/2)$, and that over $\alpha$ by making a series of substitutions. The integration over $\lambda$ (i.e. $\eta$) can then be performed, and, since we are only concerned about the limit $\kappa\rightarrow 0$, we neglect all terms that vanish in this limit. The $\rho$ integration then proceeds as above.

Having calculated the contributions near $\zeta=1$ that do not vanish as $\Delta\to 0$, we may express the divergence in terms of delta functions and plus prescriptions by comparison with the following results:
\begin{eqnarray}
\int_{1-\Delta}^1\frac{g(\zeta)}{(1-\zeta)_+}d\zeta &=& g(1)\log\Delta+\cl{O}(\Delta) \\
\int_{1-\Delta}^1\frac{g(\zeta)}{(1-\zeta)_{++}^2}d\zeta &=& -\frac{g(1)}{\Delta}-g^\prime(1)\log\Delta+g(1)+\cl{O}(\Delta) \\
\int_{1-\Delta}^1g(\zeta)\delta(1-\zeta)d\zeta &=& g(1) \\
\int_{1-\Delta}^1g(\zeta)\delta^\prime(1-\zeta)d\zeta &=& g^\prime(1)
\end{eqnarray}

Putting all this together we have
\begin{eqnarray}
C_T(\xi,\zeta) &=& \Biggl[\Gamma_T^{(0)}(\xi,\zeta)+G_T^{(0)}(\xi)\delta(1-\zeta)+H_T^{(0)}(\xi)\left(\half\delta(1-\zeta)\log(\eps\xi)-\frac{1}{(1-\zeta)_+}\right)\Biggr]\log\eps\nln
& & +\Biggl[\Gamma_T^{(1)}(\xi,\zeta)+G_T^{(1)}(\xi)\delta(1-\zeta)+H_T^{(1)}(\xi)\left(\half\delta(1-\zeta)\log(\eps\xi)-\frac{1}{(1-\zeta)_+}\right)\nln
& & +G_T^{(2)}(\xi)\delta^\prime(1-\zeta)+H_T^{(2)}(\xi)\left(\half\delta^\prime(1-\zeta)\log(\eps\xi)+\frac{1}{(1-\zeta)_{++}^2}\right)\Biggr]\eps\log\eps\\
C_L(\xi,\zeta) &=& -\frac{4(1-\xi)^2(1+2\xi^2)}{3\xi}\log\eps+\Biggl[\Gamma_L^{(1)}(\xi,\zeta)+G_L^{(1)}(\xi)\delta(1-\zeta)\nln
& & \qquad+H_L^{(1)}(\xi)\left(\half\delta(1-\zeta)\log(\eps\xi)-\frac{1}{(1-\zeta)_+}\right)\Biggr]\eps\log\eps\;,
\end{eqnarray}
where the functions $\Gamma$ do not diverge at $\zeta=1$. The explicit forms of the functions $\Gamma$, $G$ and $H$ are given in the appendix.

The terms that diverge as $\eps\to 0$ are responsible for the logarithmic scaling violations to the structure function (c.f. \cite{dmw}), while the terms proportional to $\eps\log\eps$ and $\eps\log^2\eps$ give the $1/Q^2$ power corrections. We then obtain power corrections to the generalised structure functions, given by
\beq
\delta F_i^h(x) = \frac{D_1}{Q^2}\frac{T_R C_F}{(2\pi)^2}\int_x^1\frac{d\xi}{\xi}\qtot(x/\xi)\delta C_i^h(\xi)\;,
\eeq
where
\begin{eqnarray}
\delta C_T^h(\xi) &=& \int_z^1\frac{d\zeta}{\zeta}\left[\Gamma_T^{(1)}(\xi,\zeta)-\frac{H_T^{(1)}(\xi)}{(1-\zeta)_+}+\frac{H_T^{(2)}(\xi)}{(1-\zeta)_{++}^2}\right]\Dwtd(z/\zeta)\nln
& & +\left[G_T^{(1)}(\xi)+H_T^{(1)}(\xi)\log\frac{D_2\xi^\frac{1}{2}}{Q^2}\right]\Dwtd(z)\nln
& & -\left[G_T^{(2)}(\xi)+H_T^{(2)}(\xi)\log\frac{D_2\xi^\frac{1}{2}}{Q^2}\right]\Bigl(\Dwtd(z)+z{\Dwtd}^\prime(z)\Bigr)\\
\delta C_L^h(\xi) &=& \int_z^1\frac{d\zeta}{\zeta}\left[\Gamma_L^{(1)}(\xi,\zeta)-\frac{H_L^{(1)}(\xi)}{(1-\zeta)_+}\right]\Dwtd(z/\zeta)\nln
& & +\left[G_L^{(1)}(\xi)+H_L^{(1)}(\xi)\log\frac{D_2\xi^\frac{1}{2}}{Q^2}\right]\Dwtd(z)\;,
\end{eqnarray}
and the charge-weighted sum over all quark and antiquark fragmentation functions is
\beq
\Dwtd(z) = \sum_q e_q^2 [D_q^h(z)+D_{\bar{q}}^h(z)]\;,
\eeq
while the total quark distribution in the target hadron is
\beq
\qtot(x) = \sum_q [q(x)+\bar{q}(x)]\;.
\eeq

For simplicity, let us suppose that $D_q^h(z)$ is the same for all quarks and antiquarks. This approximation is valid if there is negligible heavy flavour production and if we sum over an appropriate set of final-state hadrons. Using the definition of the fragmentation function \rf{fragdef} and using the order $\as^0$ results \rf{ord0t} - \rf{ord0b}, we find that
\beq
\qwtd(x)\delta F^h(z,x;Q^2) = (\delta F_T^h(x,z)-D_q^h(z)\delta F_T(x))+\frac{2(1-y)}{1+(1-y)^2}(\delta F_L^h(x,z)-D_q^h(z)\delta F_L(x))\;,
\eeq
where the charge-weighted quark distribution in the target hadron is
\beq
\qwtd(x) = \sum_q e_q^2 [q(x)+\bar{q}(x)]\;.
\eeq

The power corrections to the totally inclusive structure functions are given in \cite{smye,stein} as:
\beq
D_q^h(z)\delta F_i(x) = \frac{D_1}{Q^2}\frac{T_R C_F}{(2\pi)^2}\frac{\Dwtd(z)}{2}\int_x^1\frac{d\xi}{\xi}\qtot(x/\xi)\delta C_i(\xi)\;,
\eeq
where
\begin{eqnarray}
\delta C_T(\xi) &=& 2C_T^{(1)}(\xi)+2H_T^{(1)}(\xi)\log\frac{D_2\xi}{eQ^2}\\
\delta C_L(\xi) &=& 2C_L^{(1)}(\xi)+2H_L^{(1)}(\xi)\log\frac{D_2\xi}{eQ^2}\;.
\end{eqnarray}

This leads to a power correction to the fragmentation function given by
\beq
\label{res1}
\delta F^h(z,x;Q^2) = \frac{D_1}{Q^2}\frac{T_R C_F}{(2\pi)^2}\left[K_T(z,x)+\frac{2(1-y)}{1+(1-y)^2}K_L(z,x)\right]\;,
\eeq
where the transverse and longitudinal contributions are
\begin{eqnarray}
\label{res2}
K_T &=& \int_x^1\frac{d\xi}{\xi}\frac{\qtot(x/\xi)}{\qwtd(x)}\Biggl[\int_z^1\frac{d\zeta}{\zeta}\left(\Gamma_T^{(1)}(\xi,\zeta)-\frac{H_T^{(1)}(\xi)}{(1-\zeta)_+}+\frac{H_T^{(2)}(\xi)}{(1-\zeta)_{++}^2}\right)\Dwtd(z/\zeta)\nln
& & \qquad+\left(G_T^{(1)}(\xi)-C_T^{(1)}(\xi)+H_T^{(1)}(\xi)(1-\half\log\xi)\frac{}{}\right)\Dwtd(z)\nln
& & \qquad-\left(G_T^{(2)}(\xi)+H_T^{(2)}(\xi)\log\frac{D_2\xi^\frac{1}{2}}{Q^2}\right)\left(\Dwtd(z)+z{\Dwtd}^\prime(z)\right)\Biggr]\\
\label{res3}
K_L &=& \int_x^1\frac{d\xi}{\xi}\frac{\qtot(x/\xi)}{\qwtd(x)}\Biggl[\int_z^1\frac{d\zeta}{\zeta}\left(\Gamma_L^{(1)}(\xi,\zeta)-\frac{H_L^{(1)}(\xi)}{(1-\zeta)_+}\right)\Dwtd(z/\zeta)\nln
& & \qquad+\left(G_L^{(1)}(\xi)-C_L^{(1)}(\xi)+H_L^{(1)}(\xi)(1-\half\log\xi)\frac{}{}\right)\Dwtd(z)\Biggr]
\end{eqnarray}
and the explicit forms of the coefficient functions $\Gamma$, $C$, $G$ and $H$ may be found in the appendix. The quantities $D_1$ and $D_2$ are defined in equations \rf{D1def} and \rf{D2def}. Note that while there are both $1/Q^2$ and $(\log Q^2)/Q^2$ elements in the transverse contribution, the $(\log Q^2)/Q^2$ terms cancel in the longitudinal contribution.

\begin{figure}[ht]
\begin{center}
\epsfig{file=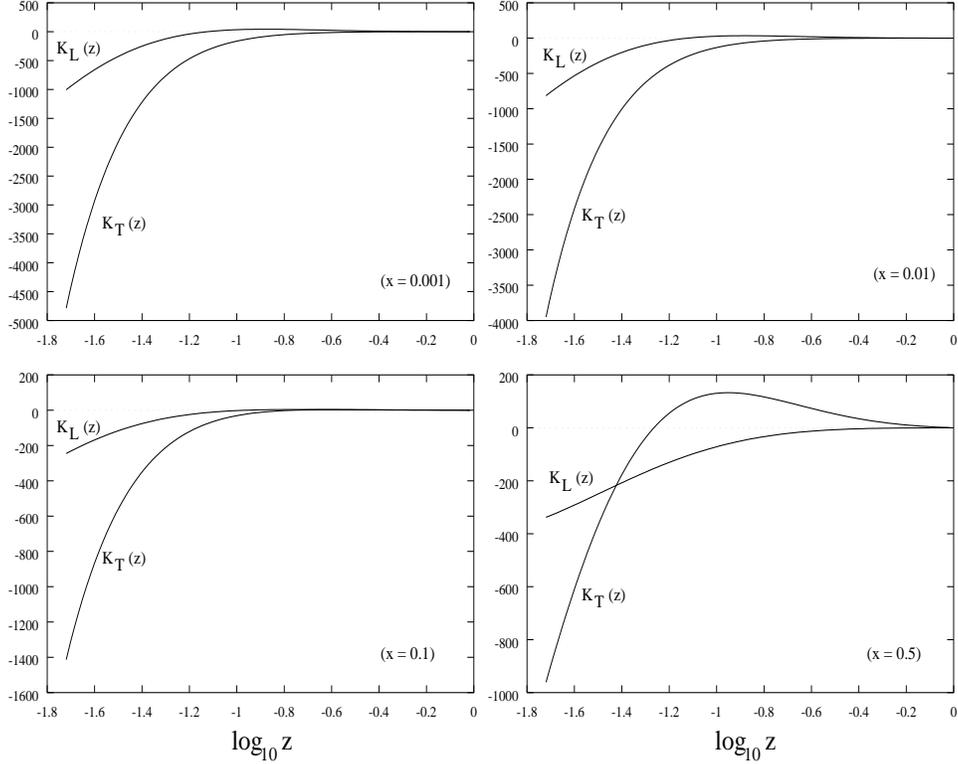, height=4.0in, width=5.0in}
\caption{\label{graph}Graphs showing $K_T(z)$ and $K_L(z)$ for various fixed values of $x$.}
\end{center}
\end{figure}

\sect{Results and Conclusions}
\label{concl}
The $1/Q^2$ power corrections arising from the diagrams shown in figure \ref{diags} are given above in equations \rf{res1}, \rf{res2} and \rf{res3}. While we do not know the values of $D_1$ and $D_2$, we can still make some qualitative predictions. Figure \ref{graph} shows plots of $K_T(z)$ and $K_L(z)$ at fixed $x$-values of 0.001, 0.01, 0.1 and 0.5. A logarithmic scale is used for the $z$ axis. These were calculated at $Q^2 = 500 \mbox{ GeV}^2$, using the corresponding MRST (central gluon) parton distributions \cite{mrst}, and ALEPH parametrisations for the light quark fragmentation functions \cite{aleph}. The value of $D_2$ was set to be $0.06 \mbox{ GeV}^2$, i.e.~approximately $\Lambda^2$, following the approach of \cite{smye} and \cite{stein}. (The qualitative behaviour of the $K_i$ does not change provided we keep $D_2 \ll Q^2$.)

The parton distributions and fragmentation functions used in the above calculation also have weak $Q^2$ dependences, behaving as $\log Q^2$, which are not shown explicitly in equations \rf{res2} and \rf{res3} but which combine with the explicit power corrections to produce additional scaling violations behaving as $(\log Q^2)/Q^2$. They are implicit within the distribution or fragmentation function and any calculation of power corrections should take the scale-evolution of these functions into account.

As can be seen from the graphs, $K_T(z)$ and $K_L(z)$ both tend to zero at large $z$, and both are divergent and negative at small $z$, with $K_T$ dominating over $K_L$. The magnitudes of $K_T$ and $K_L$ are larger at smaller $x$, although the qualitative behaviour does not change much with $x$ provided $x$ is kept reasonably small. Therefore if $D_1$ is positive, we have a negative correction to $F^h$ at small $z$. If $D_1$ is negative then the correction is positive.

We do not know the sign or magnitude of $D_1$, and there is no a priori reason why it should be either positive or negative. However, since $\cl{A}_2$ is positive, we presumably have from \rf{Aqdef} that $\delta\as(\mu^2)$ is predominantly positive. Considering the definition \rf{D1def} of $D_1$, and noting that $2\as\delta\as-(\delta\as)^2=2\asPT\delta\as+(\delta\as)^2$, we therefore might conjecture that $D_1$ would also be positive. Preliminary results from HERA may suggest a large negative correction at small $z$, lending support to this hypothesis \cite{frgres}.

\section*{Acknowledgements}
The author wishes to thank B.R.~Webber and M.~Dasgupta for advice and helpful comments.

\section*{Appendix}
\setcounter{equation}{0}
\renewcommand{\theequation}{A.\arabic{equation}}
Those functions which for brevity were not written in full in the main body of this paper are given below. Note that the functions $\Gamma$ do not diverge as $\zeta\to1$.
\begin{eqnarray}
\lefteqn{\Gamma_T^{(0)}(\xi,\zeta) = \Biggl[-\frac{4}{\xi(1-\zeta)^4}+\frac{4(1+\xi)}{\xi(1-\zeta)^3}-\frac{2(1+\xi)^2}{\xi(1-\zeta)^2}+\frac{2(1+\xi)}{1-\zeta}-\xi\Biggr]\log\left(\frac{1-\xi+\xi\zeta}{\zeta}\right)}\nln
& & +\Biggl[\frac{4(1-\xi)}{\xi(1-\zeta)^3}-\frac{2(1-\xi^2)}{\xi(1-\zeta)^2}+\frac{4(1-\xi^3)}{3\xi(1-\zeta)}\Biggr]+\frac{4-8\xi+8\xi^3-4\xi^4-3\xi^2\log\xi}{3\xi}\\
\lefteqn{\Gamma_T^{(1)}(\xi,\zeta) = \Biggl[\frac{180}{\xi(1-\zeta)^6}-\frac{192(1+\xi)}{\xi(1-\zeta)^5}+\frac{6(9+34\xi+9\xi^2)}{\xi(1-\zeta)^4}-\frac{4(1+14\xi+14\xi^2+\xi^3)}{\xi(1-\zeta)^3}}\nln
& & +\frac{4+19\xi+4\xi^2}{(1-\zeta)^2}-\frac{2\xi(1+\xi)}{1-\zeta}\Biggr]\log\left(\frac{1-\xi+\xi\zeta}{\zeta}\right)+\Biggl[-\frac{180(1-\xi)}{\xi(1-\zeta)^5}+\frac{102(1-\xi^2)}{\xi(1-\zeta)^4}\nln
& & -\frac{18(1+3\xi-3\xi^2-\xi^3)}{\xi(1-\zeta)^3}-\frac{2(2-7\xi+7\xi^3-2\xi^4)}{\xi(1-\zeta)^2}-\frac{(1-\xi)(4+7\xi^2+4\xi^4)}{\xi(1-\zeta)}\Biggr]\nln
& & +\frac{2\xi^3(1-2\xi+2\xi^2)}{1-\xi+\xi\zeta}-\frac{2(2-2\xi+\xi^2)}{\xi\zeta}\\
\lefteqn{\Gamma_L^{(1)}(\xi,\zeta) = \Biggl[\frac{144}{\xi(1-\zeta)^6}-\frac{192(1+\xi)}{\xi(1-\zeta)^5}+\frac{8(7+30\xi+9\xi^2)}{\xi(1-\zeta)^4}-\frac{8(8+10\xi+\xi^2)}{(1-\zeta)^3}}\nln
& & +\frac{4(5\xi+2\xi^2)}{(1-\zeta)^2}-\frac{4\xi^2}{1-\zeta}\Biggr]\log\left(\frac{1-\xi+\xi\zeta}{\zeta}\right)+\Biggl[-\frac{144(1-\xi)}{\xi(1-\zeta)^5}+\frac{120(1-\xi^2)}{\xi(1-\zeta)^4}\nln
& & -\frac{8(1+11\xi-9\xi^2-3\xi^3)}{\xi(1-\zeta)^3}+\frac{8(1+\xi-2\xi^2)}{(1-\zeta)^2}+\frac{4(2-15\xi^2+10\xi^3+3\xi^5)}{15\xi(1-\zeta)}\Biggr]
\end{eqnarray}
\begin{eqnarray}
C_T^{(1)} &=& \frac{2+25\xi^2-25\xi^3-2\xi^5+15\xi^2\log\xi+15\xi^3\log\xi}{5\xi}\\
C_L^{(1)} &=& -\frac{4(17+75\xi^2-125\xi^3+33\xi^5+30\log\xi+75\xi^3\log\xi-45\xi^5\log\xi)}{225\xi}\\
G_T^{(0)} &=& -\frac{26-90\xi+90\xi^2-26\xi^3+12\log\xi-9\xi\log\xi-9\xi^2\log\xi+12\xi^3\log\xi}{18\xi}\\
G_T^{(1)} &=& \frac{6+20\xi+90\xi^2-90\xi^3-20\xi^4-6\xi^5+45\xi^2\log\xi+75\xi^3\log\xi}{30\xi}\\
G_T^{(2)} &=& \frac{7-20\xi+20\xi^3-7\xi^4-24\xi^2\log\xi}{12\xi}\\
G_L^{(1)} &=& -2(92+150\xi^2-425\xi^3+183\xi^5+30\log\xi-225\xi^2\log\xi\nln
& & \qquad+150\xi^3\log\xi+45\xi^5\log\xi)/(225\xi)\frac{}{}\\
H_T^{(0)} &=& \frac{4+3\xi-3\xi^2-4\xi^3+6\xi\log\xi+6\xi^2\log\xi}{3\xi}\\
H_T^{(1)} &=& -\xi(5-5\xi+2\log\xi+2\xi\log\xi)\frac{}{}\\
H_T^{(2)} &=& \xi\log\xi\frac{}{}\\
H_L^{(1)} &=& \frac{4(2-15\xi^2+10\xi^3+3\xi^5-15\xi^3\log\xi)}{15\xi}\;.
\end{eqnarray}


\begin{thebibliography}{50}

\bibitem{expfrgfns}
ZEUS Collaboration, M.~Derrick et al.: \zphysc{67} (1995) 93, (hep-ex 9501012).\\
H1 Collaboration, S.~Aid et al.: \nphysb{445} (1995) 3, (hep-ex 9505003).\\
C.~Adloff et al.: \nphysb{504} (1997) 3, (hep-ex 9707005).

\bibitem{renlons}
For reviews and classic references see V.I.~Zakharov: \nphysb{385} (1992) 452, and A.H.~Mueller, in {\em QCD 20 Years Later}, vol.~{\bf 1} (World Scientific, Singapore, 1993). \\
M.~Beneke, V.M.~Braun and V.I.~Zakharov: \prlett{73} (1994) 3058, (hep-ph 9405304). \\
P.~Ball, M.~Beneke and V.M.~Braun: \nphysb{452} (1995) 563, (hep-ph 9502300). \\
M.~Beneke and V.M.~Braun: \plettb{348} (1995) 513, (hep-ph 9411229). \\
M.~Beneke and V.M.~Braun: \nphysb{454} (1995) 253, (hep-ph 9506452). \\
M.~Beneke: CERN-TH/98-233, (hep-ph 9807443). \\
M.~Neubert: \phrevd{51} (1995) 5924, (hep-ph 9412265). \\
Yu.L.~Dokshitzer and N.G.~Uraltsev: \plettb{380} (1996) 141, (hep-ph 9512407). \\
G.~Grunberg: CPTH-PC463-0896, (hep-ph 9608375). \\
G.~Grunberg: Ecole Polytechnique preprint CPTH-S505-0597, (hep-ph 9705290).


\bibitem{dmw}
Yu.L.~Dokshitzer, G.~Marchesini and B.R.~Webber: \nphysb{469} (1996) 93, (hep-ph 9512336).

\bibitem{asmodel}
Yu.L.~Dokshitzer, V.A.~Khoze and S.I.~Troyan: \phrevd{53} (1996) 89, (hep-ph 9506425). \\
D.V.~Shirkov and I.L.~Solovstov: \prlett{79} (1997) 1209, (hep-ph 9704333). \\
G.~Grunberg: \plettb{372} (1996) 121, (hep-ph 9512203). \\
D.V.~Shirkov and I.L.~Solovtsov: \jinrrc{2 [76]} (1996) 5, (hep-ph 9604363). \\
D.V.~Shirkov: \npproc{64} (1998) 106, (hep-ph 9708480). \\
A.I.~Alekseev and B.A.~Arbuzov: \mpleta{13} (1998) 1747, (hep-ph 9704228). \\
B.R.~Webber: \jhephy{10} (1998) 012, (hep-ph 9805484).

\bibitem{eefrag}
M.~Dasgupta and B.R.~Webber: \nphysb{484} (1997) 247, (hep-ph 9608394). \\
P.~Nason and B.R.~Webber: \plettb{395} (1997) 355, (hep-ph 9612353). \\
M.~Beneke, V.M.~Braun and L.~Magnea, \nphysb{497} (1997) 297, (hep-ph 9701309).

\bibitem{eeevsh}
B.R.~Webber: \plettb{339} (1994) 147.
Yu.L.~Dokshitzer and B.R.~Webber: \plettb{352} (1995) 451, (hep-ph 9504219). \\
G.P.~Korchemsky and G.~Sterman: \nphysb{437} (1995) 415, (hep-ph 9411211). \\
G.P.~Korchemsky and G.~Sterman: in the Proceedings of the 30th Rencontres de Moriond, Meribel les Allues, France, March 1995, (hep-ph 9505391). \\
G.P.~Korchemsky, G.~Oderda and G.~Sterman: to appear in the Proceedings of the 5th International Workshop on DIS and QCD, (hep-ph 9708346). \\
R.~Akhoury and V.I.~Zakharov: \plettb{357} (1995) 646, (hep-ph 9504248). \\
R.~Akhoury and V.I.~Zakharov: \nphysb{465} (1996) 295, (hep-ph 9507253). \\
P.~Nason and M.H.~Seymour: \nphysb{454} (1995) 291, (hep-ph 9506317). \\
Yu.L.~Dokshitzer and B.R.~Webber: \plettb{404} (1997) 321, (hep-ph 9704298).

\bibitem{dwdisstr}
M.~Dasgupta and B.R.~Webber: \plettb{382} (1996) 273, (hep-ph 9604388).

\bibitem{disstr}
E.~Stein, M.~Meyer-Hermann, L.~Mankiewicz and A.~Sch\"{a}fer: \plettb{376} (1996) 177, (hep-ph 9601356). \\
M.~Meyer-Hermann, M.~Maul, L.~Mankiewicz, E.~Stein and A.~Sch\"{a}fer: \plettb{383} (1996) 463, (hep-ph 9605229); ibid.~{\bf 393} (1997) 487 (E). \\
M.~Maul, E.~Stein, A.~Sch\"{a}fer and L.~Mankiewicz, \plettb{401} (1997) 100, (hep-ph 9612300). \\
M.~Maul, E.~Stein, L.~Mankiewicz, M.~Meyer-Hermann and A.~Sch\"{a}fer, (hep-ph 9710392). \\
M.~Meyer-Hermann and A.~Sch\"{a}fer, (hep-ph 9709349).

\bibitem{disfrg}
M.~Dasgupta, G.E.~Smye and B.R.~Webber: \jhephy{04} (1998) 017, (hep-ph 9803382).

\bibitem{disev}
M.~Dasgupta and B.R.~Webber: \ephysj{C1} (1998) 539, (hep-ph 9704297). \\
M.~Dasgupta and B.R.~Webber: Cambridge preprint Cavendish-HEP-98/02, (hep-ph 9809247).

\bibitem{milan}
Yu.L.~Dokshitzer, A.~Lucenti, G.~Marchesini and G.P.~Salam: \nphysb{511} (1998) 396, (hep-ph 9707352). \\
Yu.L.~Dokshitzer, A.~Lucenti, G.~Marchesini and G.P.~Salam: \jhephy{05} (1998) 003, (hep-ph 9802381). \\
G.P.~Salam: Milan preprint IFUM-623-FT, (hep-ph 9805323).

\bibitem{hautmann}
F.~Hautmann: \prlett{80} (1998) 3198, (hep-ph 9710256).

\bibitem{smye}
G.E.~Smye: Cavendish-HEP-98/14, (hep-ph 9810292).

\bibitem{stein}
E.~Stein, M.~Maul, L.~Mankiewicz and A.~Sch\"{a}fer: Torino preprint DFTT 13/97, to appear in \nphysb{}, (hep-ph 9803342).

\bibitem{bframe}
R.D.~Peccei and R.~R\"{u}ckl: \nphysb{162} (1980) 125. \\
K.H.~Streng, T.F.~Walsh and P.M.~Zerwas: \zphysc{2} (1979) 237. \\
L.V.~Gribov, Yu.L.~Dokshitzer, S.I.~Troyan and V.A.~Khoze: \sovphy{68} (1988) 1303.

\bibitem{mrst}
A.D.~Martin, R.G.~Roberts, W.J.~Stirling and R.S.~Thorne: \ephysj{C1} (1998) 463, (hep-ph 9803445).

\bibitem{aleph}
ALEPH Collaboration, D.~Buskulic et al.: \plettb{357} (1995) 487, \plettb{364} (1995) 247 (E).

\bibitem{frgres}
H1 Collaboration: submission 531 to the 29th International Conference on High-Energy Physics, Vancouver, 1998. \\
ZEUS Collaboration: submission 809 to the 29th International Conference on High-Energy Physics, Vancouver, 1998.

\end{thebibliography}
\end{document}